\documentclass[twocolumn]{aastex631}
\usepackage[utf8]{inputenc}
\usepackage{natbib}
\setcitestyle{round}
\usepackage{amsmath,amssymb}
\usepackage{afterpage}
\usepackage{hyperref}

\usepackage{xcolor}
\usepackage[caption=false]{subfig}
\usepackage{soul}
\usepackage{enumitem}
\usepackage{xcolor}
\usepackage{graphicx}

\begin{document}

\title{Dynamical Interactions and Habitability in the TOI-700 Multi-Planet System}

\author[0009-0009-2277-799X]{Coleman Nelson}
\affiliation{University of Wisconsin-Madison Department of Astronomy, 475 N. Charter St. Madison, WI 53706, USA}

\author[0000-0002-7733-4522]{Juliette Becker}
\affiliation{University of Wisconsin-Madison Department of Astronomy, 475 N. Charter St. Madison, WI 53706, USA}

\begin{abstract}
    The discovery of a second earth sized planet (TOI-700e) interior to the habitable candidate TOI-700d has prompted further research into this system, as the additional planet makes the TOI-700 system a tightly packed multi-planet system with multiple planets in the habitable zone, like TRAPPIST-1. In this work, we use the planetary evolution code \texttt{VPLanet} to assess the potential habitability of TOI-700d and TOI-700e. We first examine their orbital dynamics to evaluate the influence of planet–planet interactions on {the planet spin, obliquity, and} eccentricity. We then investigate whether {these interactions are sufficient to cause either TOI-700d or e to be perturbed out of a habitable state, and whether we expect} either planet could retain surface oceans over Gyr timescales. Together, these analyses allow us to assess the long-term habitability prospects of both TOI-700d and TOI-700e. We find that multi-planet interactions in the TOI-700 system do not prevent either planet from potentially retaining habitable conditions; {however, we find that TOI-700e is located very near the boundary of the tidally locked habitable zone \citep{Kopparapu2017}, suggesting further work is needed to determine whether it is truly habitable}.
\end{abstract}

\section{Introduction}
The data-based study of planetary habitability has emerged as a field over the past decade, as the exoplanet census has expanded due to work by various observational surveys \citep{Borucki2010, Howell2014, Guerrero2021} to include a growing number of potentially habitable planets \citep{Kane2016}, particularly orbiting M-dwarfs \citep{Shields2016}.
These new candidates have enabled efforts to understand the nuances of habitability in specific exoplanet systems \citep[e.g.,][]{Bolmont2014, Daspute2025} and find and characterize possible biosignatures{ \citep[e.g.,][]{Rotman2023, Eager-Nash2024, Schwieterman2024}. }

The effective temperature of a planet, often used as a proxy for habitability due to its role in determining whether liquid water can exist on the surface, is a widely adopted metric for classifying exoplanets as habitable or not {\citep{Manabe1964}}. 
{The exact planetary properties, including the} atmospheric composition \citep{Kasting1993, Kopparapu2013}, surface water temperature \citep{Mitchell2025}, obliquity \citep{Lerner2025}, eccentricity \citep{Way2023, Liu2024}, and multi-body interactions \citep{Quarles2019}, can significantly impact the chances for water to exist in liquid form on a planet's surface. {In particular, these factors all impact the heat redistribution, meaning that the same computed surface temperature could correspond to very different habitability prospects on planets with different properties \citep{Lefevre&Turbet2021}.} 
As a result, while the computed planetary effective temperature is a good bulk classifier for identifying habitable candidates, more detailed analyses are needed to understand the true likelihood of habitability for individual planets and planetary systems.
Such detailed analyses for individual systems have made it evident that habitability must also be considered as time-dependent, since the orbital dynamics of some planets cause them to periodically enter and exit their habitable zones \citep{Damasso2022}.

\begin{figure*}
    \centering
    \includegraphics[width=1\linewidth]{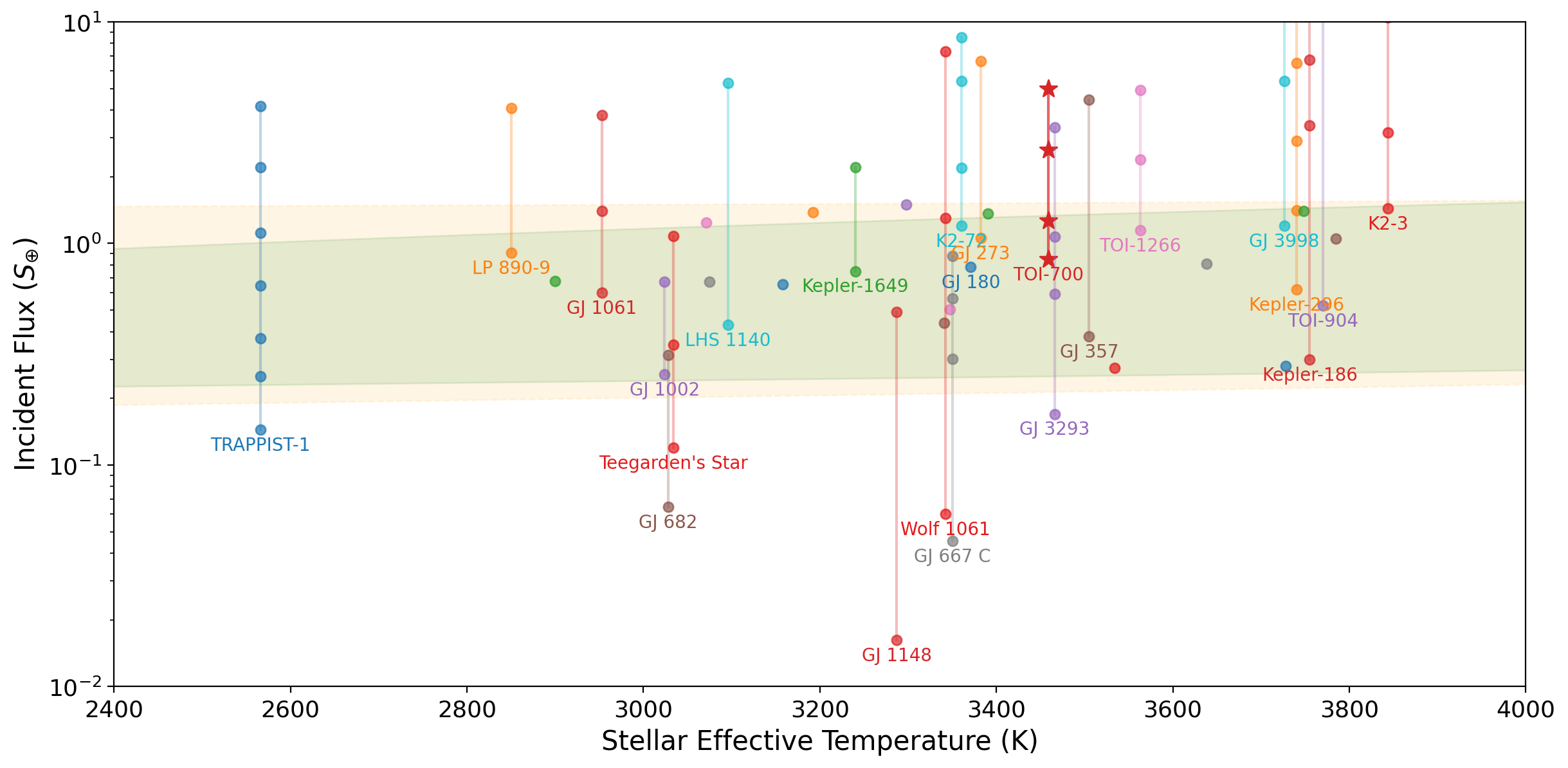}
    \caption{Incident flux in $S_\oplus$ received by planets orbiting M-dwarfs in systems where at least one planet resides in the habitable zone as a function of their host star's effective temperature. Each vertical line connects planets within the same system. Systems with at least one planet in the optimistic habitable zone (shaded orange) or conservative habitable zone (shaded green) are included. Multi-planet systems are labeled with the system name. {The optimistic habitable zone (orange shading) is bounded by the Recent Venus and Early Mars {1D} limits from \citet{Kopparapu2014}. The conservative habitable zone (green shading) uses a more restrictive definition: its inner boundary is set by the stellar insolation shown by {the 3D models of} \citet{Kopparapu2017} to trigger a runaway greenhouse effect for tidally locked planets around M-dwarfs, while its outer boundary follows the maximum greenhouse limit assumed by \citet{Kopparapu2014}.} {While the location of the inner edge was computed using 3D models, the outer edge relies solely on 1D models and may therefore be systematically offset from its true position \citep[e.g.,]{Leconte2013}.}  Tightly packed multi-planet systems with multiple planets in the habitable zone include TRAPPIST-1, GJ 3293, and TOI-700. Data obtained 6/7/2025 from the IPAC Exoplanet Archive \citep{Christiansen2025}.  }    \label{fig:habzone}
\end{figure*}

One particularly interesting target for such habitability studies is the TOI-700 system.  
TOI-700 is a compact, multi-planet system orbiting an early M dwarf, discovered using data from the Transiting Exoplanet Survey Satellite (TESS) \citep{Gilbert2020, Rodriguez2020}. Its three initially identified planets (TOI-700 b, c, and d) included one potentially habitable planet, as TOI-700d resides within the conservative habitable zone \citep{Kopparapu2013} of the host star. TOI-700d is likely a rocky planet, and early modeling work suggested that it could retain surface water in liquid or frozen form, depending on atmospheric conditions \citep{Suissa2020, Dong2020, Cohen2020, Bonney2022}. In 2023, the discovery of a fourth planet, TOI-700e, an Earth-sized planet orbiting between TOI-700 c and TOI-700d, further heightened interest in this system \citep{Gilbert2023}. TOI-700e receives higher insolation than TOI-700d and lies in the optimistic habitable zone, raising the possibility that two adjacent planets in the system might be capable of supporting liquid water.

The tight orbital spacing of the TOI-700 system (illustrated in Figure~\ref{fig:habzone}) places it in a class of dynamically compact M dwarf systems, similar to TRAPPIST-1, where planet–planet interactions may play an important role in shaping orbital and climatic evolution.
In the well-studied TRAPPIST-1 system, multi-body interactions such as planet-planet tides \citep{Wright2018, Hay2019} are expected to contribute to the climate states and habitability prospects of the planets. 
In the TOI-700 system, the proximity of the two potentially habitable planets TOI-700~d and TOI-700~e, in particular, suggests that planet-planet interactions could influence climate-controlling parameters such as eccentricity and obliquity. 

{The discovery of TOI-700e in 2023 \citep{Gilbert2023}, orbiting between TOI-700c and TOI-700d, revealed that the habitable-zone planet TOI-700d is more closely spaced with its neighboring planets than was initially realized at the time of its discovery in 2020 \citep{Gilbert2020}. This begs the question: does this additional planet put the habitability prospects of TOI-700d at risk?}
In this work, we assess the potential habitability of the planets in the TOI-700 system, with a particular focus on how planet–planet interactions among the closely spaced, potentially habitable planets may impact the system's long-term habitability. 
In Section \ref{sec:methods}, we present the constraints on planetary habitability derived from our \texttt{VPLanet} simulation results: Section \ref{sec:obl} presents an analysis of the rotational evolution of the planets using \texttt{distorb}, \texttt{distrot}, and \texttt{eqtide}; Section \ref{subsec:ecc} presents a secular analysis of the evolving planetary flux using \texttt{distorb}, \texttt{distrot}, and \texttt{stellar}; and Section \ref{subsec:water} presents the results of our analysis of water loss using \texttt{atmesc} and \texttt{stellar}. Section \ref{sec:discuss} discusses the context and implications of our findings. Finally, we conclude in Section \ref{sec:conclude} with a summary of our results.

\section{Modeling Planetary Habitability in TOI-700} 
\label{sec:methods}
To assess the degree to which multi-planet interactions affect the habitability states of the planets in the TOI-700 system, we use the modular planetary habitability code \texttt{VPLanet} \citep{Barnes2020}. 
VPLanet is a multipurpose planetary evolution code that has many separate modules that model each of the individual processes that occur during planetary evolution. 
These modules can be combined to study more complex effects \citep[e.g.,][]{doAmaral2022, Wilhelm2022, Livesey2024, Fromont2024, Barnes2025}.

In this work, we use several modules that track parameters that affect the potential habitability of the TOI-700 system. 
{The \texttt{stellar} module models the evolution of fundamental properties of low-mass stars ($M_*\leq 1.4 M_{\odot}$), using interpolation off the grids of MESA stellar evolution models computed by \citet{Baraffe2015}.}
The \texttt{distorb} module simulates orbital evolution by using the secular disturbing function \citep{MD99, Ellis2000}, which {computes the} stability of the planets and determines the impact of planet-planet interactions in such a compact system. 
{The \texttt{distrot} module tracks the planetary rotation and obliquity using a similar disturbing function derived by \citet{Kinoshita1975}. This module tracks the rotational evolution of the planet from orbits and the stellar torque.} 
Finally, we use the \texttt{atmesc} module to track the rate at which gas escapes from the atmosphere of planets d and e. { The \texttt{atmesc} module models the escape of planetary atmospheres by using simple parametric energy-limited and diffusion-limited prescriptions.}
{More information about the modules used in our analysis and their specific implementations can be found in the Appendices of \citet{Barnes2020}. }
 While we include all four known planets in each simulations, we focus on the parameter variations for the two habitable planets, TOI-700e and TOI-700d.   

\begin{table*}
    \centering
    \begin{tabular}{lccc}
        \hline
        Parameter & Description (Units) & Adopted Value  & Ref. \\
        \hline
        $M_*$  & Mass ($M_{\odot})$ & $0.415_{-0.020}^{+0.021}$ & \citet{Gilbert2023} \\
        $R_*$  & Radius ($R_\odot$) & $0.42_{-0.015}^{+0.017}$ & \citet{Gilbert2023}  \\
        $L_*$ & Luminosity ($\L_{\odot}$) & $0.02229_{-0.0023}^{+0.0026}$ & \citet{Gilbert2023}  \\
        $T_{eff}$ & Effective Temperature (K) & 3359 $\pm 65$ & \citet{Gilbert2023} \\
        $[Fe/H]$ & Metallicity (dex) & -0.07 $\pm 0.11$ & \citet{Gilbert2023}  \\
        $P_{\rm{rot}}$ & Rotation Period (days) & 54 $\pm 0.8$ & \citet{Gilbert2020} \\
        $k_{2,*}$ & Love number  & 0.3 & Assumed \\
        $Q_{*}$ & Tidal Quality Factor  & $10^7$ & Assumed \\
        \hline\hline

\end{tabular}
        \caption{Stellar parameters for the M dwarf TOI-700 used in this work. {  }    } 
        \label{tab: stellar}
\end{table*}

\begin{table*}
    \centering
    \begin{tabular}{lcccc}
        \hline
        Parameter & Description (Units) & Adopted Value & Ref. & Comments \\
        \hline
        \multicolumn{5}{l}{\textbf{TOI-700 b}} \\
        $P$ & Period (days) & $9.977219_{-0.000038}^{+0.000041}$ & \citet{Gilbert2023} &  \\
        $R_P$ & Radius ($R_\oplus$) & $0.914_{-0.049}^{+0.053}$ & \citet{Gilbert2023} &  \\
        $a$ & Semimajor axis (AU) & $0.0677 \pm{0.0011}$ & \citet{Gilbert2023} &  \\
        $e$ & Eccentricity & $0.075_{-0.054}^{+0.093}$ & \citet{Gilbert2023} &  \\
        $S$ & Insolation flux (So) & $4.98_{-0.58}^{+0.50}$ & \citet{Gilbert2023} &  \\
        $M_P$ & Planetary Mass ($M_\oplus$) & $\sim 1$ & \citet{Weiss2014} & Inferred$^{1}$  \\
        $k_{2,p}$ & Love number  & 0.3 & & Assumed \\
        $Q_p$ & Tidal Quality Factor & 100 &  & Assumed  \\

        \\

        \multicolumn{5}{l}{\textbf{TOI-700 c}} \\
        $P$ & Period (days) & $16.051137 \pm 0.000020$ & \citet{Gilbert2023} &  \\
        $R_P$ & Radius ($R_\oplus$) & $2.60_{-0.13}^{+0.14}$ & \citet{Gilbert2023} &  \\
        $a$ & Semimajor axis (AU) & $0.0929 \pm 0.0015$ & \citet{Gilbert2023} &  \\
        $e$ & Eccentricity & $0.068_{-0.049}^{+0.070}$ & \citet{Gilbert2023} &  \\
        $S$ & Insolation flux (So) & $2.64_{-0.31}^{+0.26}$ & \citet{Gilbert2023} &  \\
        $M_P$ & Planetary Mass ($M_\oplus$) & $\sim6.11$ & \citet{Weiss2014} & Inferred$^{1}$  \\
        $k_{2,p}$ & Love number  & 0.3 & & Assumed \\
        $Q_p$ & Tidal Quality Factor & 1000 &  & Assumed  \\
\\
        \multicolumn{5}{l}{\textbf{TOI-700e}} \\
        $P$ & Period (days) & $27.80978_{-0.00040}^{+0.00046}$ & \citet{Gilbert2023} &  \\
        $R_P$ & Radius ($R_\oplus$) & $0.953_{-0.075}^{+0.089}$ & \citet{Gilbert2023} &  \\
        $a$ & Semimajor axis (AU) & $0.1340 \pm 0.0022$ & \citet{Gilbert2023} &  \\
        $e$ & Eccentricity & $0.059_{-0.042}^{+0.057}$ & \citet{Gilbert2023} &  \\
        $S$ & Insolation flux ($S_\oplus$) & $1.27_{-0.15}^{+0.13}$ & \citet{Gilbert2023} &  \\
        $M_P$ & Planetary Mass ($M_\oplus$) & $\sim0.86$ & \citet{Weiss2014} & Inferred$^{1}$  \\
        $k_{2,p}$ & Love number  & 0.3 & & Assumed \\
        $Q_p$ & Tidal Quality Factor & 100 &  & Assumed  \\
\\
        \multicolumn{5}{l}{\textbf{TOI-700d}} \\
        $P$ & Period (days) & $37.42396_{-0.00035}^{+0.00039}$ & \citet{Gilbert2023} &  \\
        $R_P$ & Radius ($R_\oplus$) & $1.073_{-0.054}^{+0.059}$ & \citet{Gilbert2023} &  \\
        $a$ & Semimajor axis (AU) & $0.1633 \pm 0.0027$ & \citet{Gilbert2023} &  \\
        $e$ & Eccentricity & $0.042_{-0.030}^{+0.045}$ & \citet{Gilbert2023} &   \\
        $S$ & Insolation flux ($S_\oplus$) & $0.85_{-0.10}^{+0.09}$ & \citet{Gilbert2023} &  \\
        $M_P$ & Planetary Mass ($M_\oplus$) & $\sim1.69$ & \citet{Weiss2014} & Inferred$^{1}$ \\
        $k_{2,p}$ & Love number  & 0.3 & & Assumed \\
        $Q_p$ & Tidal Quality Factor & 100 &  & Assumed  \\

        \hline\hline
    \end{tabular}
    \caption{Median values and 68\% confidence intervals{, when available, for the planetary parameters of TOI-700. Most physical parameters are taken directly from the global solution from \citet{Gilbert2023}. For parameters needed for our analysis where measurements do not exist, we provide our assumed values. $^{1}$: Planet masses have not yet been measured at the time this work was completed, and so are estimated using the mass-radius relation of \citet{Weiss2014}. Insolation flux for each planet is reported in solirad (So), defined as the nominal total solar irradiance \citep{Mamajek2025}.}}
    \label{tab:bigtable}
\end{table*}

\begin{figure}
    \centering
    \includegraphics[width=1\linewidth]{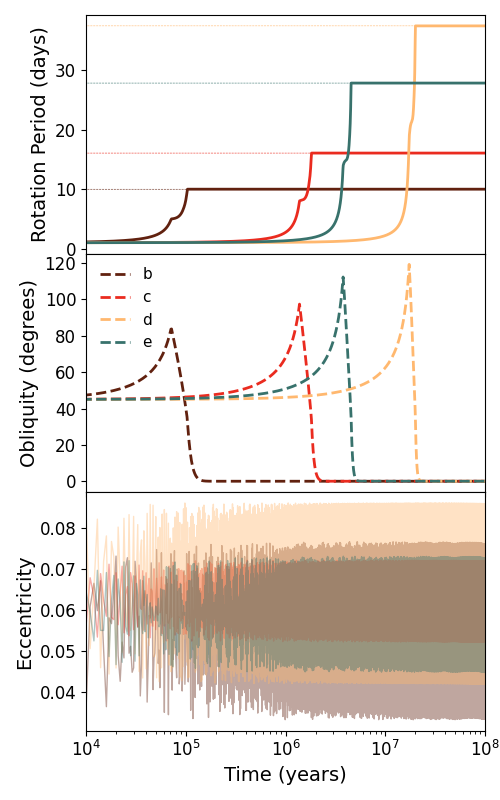}
    \caption{{Time evolution of planetary spin–orbit properties for planets TOI-700 b–e. \textit{Top panel}: Rotation period as a function of time, with horizontal dotted lines indicating the present-day observed orbital periods for each planet. When the rotation and orbital periods for a planet are equal, the planet is considered tidally locked. \textit{Middle panel:} Obliquity evolution, showing growth toward high obliquity followed by rapid damping to near-zero as each planet transitions into an equilibrium state (which occurs sooner for planets with shorter orbital periods). \textit{Bottom panel:} Orbital eccentricity evolution, illustrating secular evolution of eccentricity.}}
    \label{fig:obl_evolution}
\end{figure}

\subsection{Tidal Evolution and Obliquity}
\label{sec:obl}
{As pointed out in \citet{Gilbert2020}, it is likely that the planets in this system will be tidally locked, defined as a state where the orbital period of a planet is equal to its spin period. The tidal locking timescale can be written as \citep{Gladman1996}:}

\begin{equation}
\tau_{\mathrm{lock}}=\frac{w\,a^{6} C Q}{3G\,M_{\ast}^{2}k_{2}R_{p}^{5}}\,
\label{eq:tlock}
\end{equation}
{when $\omega$ is the initial planet spin frequency, $a$ is the semi-major axis, $C \propto M_{p}R_{p}^2$ the planet's moment of inertia, $Q$ the tidal quality factor of the planet, $G$ the gravitational constant, $M_\ast$ the stellar mass, $k_2$ the planetary love number, and $R_p$ the planetary radius. Using the best fit values from Table \ref{tab:bigtable} yields a tidal locking timescale of $\sim$10 Myr for TOI-700d, the outermost known planet. This suggests that all the planets in the TOI-700 system are likely tidally locked, since $\tau_{\mathrm{lock}} \propto a^6.$}
 
{To model the spin and orbital evolution of the planets in the TOI-700 system, we use the VPLanet modules \texttt{eqtide}, \texttt{distorb}, and \texttt{distrot}. Tidal dissipation is treated using the \texttt{eqtide} implementation of the constant phase lag equilibrium tide model, adopting Love numbers $k_2 = 0.3$ for both the star and planets, which corresponds to a fully convective body modeled as an $n = 3/2$ polytrope \citep[e.g.,][]{Ragozzine2009, Feiden2011}. Secular spin–orbit coupling and orbital evolution are computed using the second-order secular disturbing function \citep{MD99} implemented in the \texttt{distorb} and \texttt{distrot} modules. 
We assume planetary tidal quality factors of $Q = 100$ for planets b, d, and e, while the larger-radius planet c is assigned a higher value of $Q = 1000$ to reflect its likely more substantial envelope and subsequent weaker tidal dissipation. The tidal quality factor $Q$ is likely to range between $10^6- 10^8$ for M dwarfs \citep{Ilic2024}, and so the host star is assumed to have $Q_\star = 10^7$ and responds to tides raised by all planets. Orbital elements, spin rates, and obliquities are evolved self-consistently with an adaptive timestep from an initially slightly eccentric (best fit eccentricities as shown in Table \ref{tab:bigtable}, all lower that 0.1), moderately high obliquity (45 degrees for all planets), and fast initial rotation periods of 1 day. Numerical simulations of planet formation show that young planets are expected to rotate rapidly as a consequence of planetesimal accretion and can also acquire substantial obliquities through stochastic giant impacts \citep{Miguel2010}. We therefore adopt dynamically active initial conditions in order to assess the timescales over which tidal dissipation damps the planets’ primordial spins and obliquities.}

{The results of this set of simulations is shown in Figure \ref{fig:obl_evolution}. In the top panel, all planets exhibit spin-down driven by tidal dissipation, with their rotation periods converging toward their respective orbital periods (marked as dashed lines). This marks the onset of synchronous rotation. As expected from Equation \ref{eq:tlock}, the innermost planets reach spin–orbit synchronization first. The middle panel shows the obliquity evolution, with each planet initially experiencing excitation to high obliquity followed by rapid damping toward near-zero values as tidal dissipation drives the system toward an equilibrium spin state. The range of eccentricities in the bottom panel show consistent values with the current-day observed values. Despite the fact that tidal effects would tend to decrease eccentricities with time, planet–planet interactions are strong enough that the eccentricities are sustained. }

{We also ran two additional suites of integrations with lower and upper limits on the tidal quality factors $Q$ for each of the planets. The chosen lower limits were $Q=10$ for TOI-700 b, d and e and $Q=100$ for TOI-700 c, and the upper limits were $Q=300$ for TOI-700 b, d and e and $Q=1000$ for TOI-700 c. Even for the least dissipative cases, tidal locking occurred within 100 Myr for the outer planet TOI-700d, making it very likely that all the planets in the system are tidally locked. }

{It is important to note that for planets with non-zero orbital eccentricity, the state of $P_{\rm{orb}} = P_{\rm{spin}}$ does not necessarily mean that the planet will have a fixed dayside and nightside. In eccentric orbits, the equilibrium state corresponds to pseudo-synchronous rotation \citep{Selsis2007, Heller2011}, in which the instantaneous angular velocity varies over the orbit. However, the planets in the TOI-700 system maintain relatively low eccentricities throughout their evolution, and so the difference between true synchronous and pseudo-synchronous rotation is small. As a result, it is likely that all the known planets in the TOI-700 systems have persistent day–night asymmetries, unless there is significant planetary obliquity - and even then, some portion of the planet might still be eternally dark. } 
 
\subsection{Eccentricity Evolution and Insolation}
\label{subsec:ecc}

\begin{figure}
    \centering
    \includegraphics[width=1\linewidth]{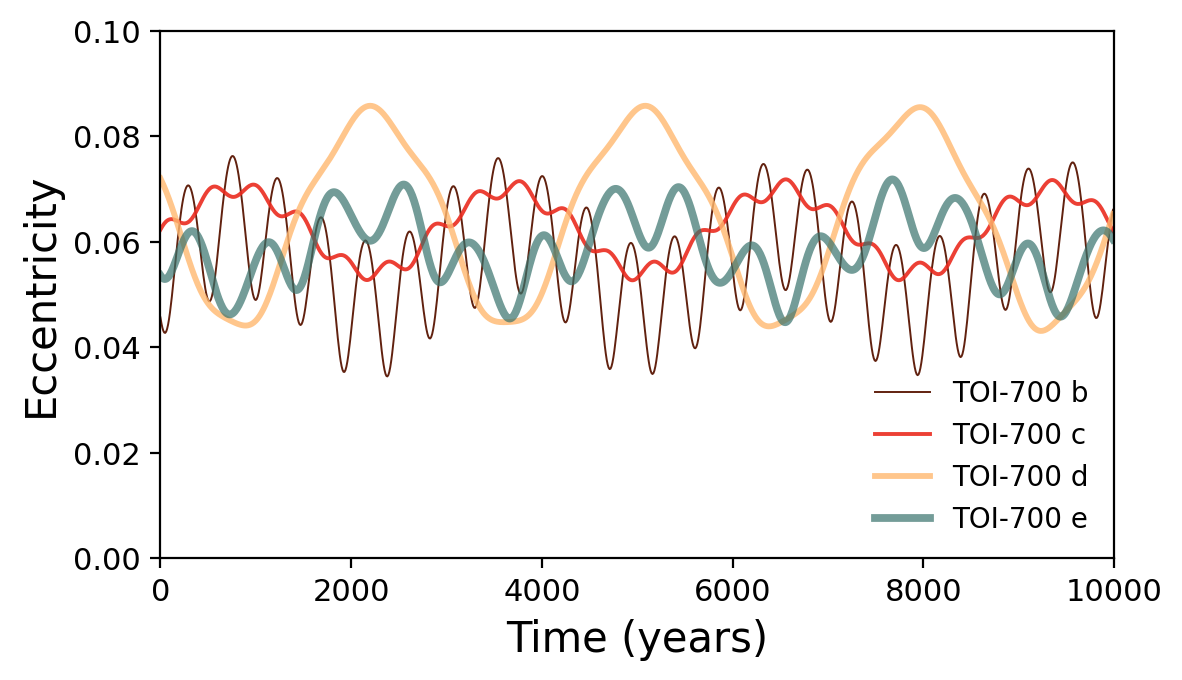}
    \caption{Eccentricity evolution {of the four planets in the TOI-700 system over 10,000 years of the integration after 100 Myr of evolution. This plot is a zoom-in of a small range of bottom panel of Figure \ref{fig:obl_evolution}.}. Additionally, all planets have low {but non-zero} eccentricity {$0.03 < e<0.1$} over their entire evolution.}
    \label{fig:ecc_e_d}
\end{figure}

A factor that affects the habitability of planets, particularly those in tightly packed multi-planet systems, is planet-planet interactions. 
Secular interactions between planets naturally cause their orbital elements, including eccentricity, to evolve over time. If the planets are particularly closely spaced and these effects are strong, it can lead to variations in the planets’ levels of irradiation over time, potentially affecting a planets' prospects for habitability. 
{In this section, we consider the effect of secular oscillations in eccentricity on the expected insolation levels of the TOI-700 system, after the tidal evolution modeled in the previous section has proceeded.}

{In Figure \ref{fig:ecc_e_d}, we show the eccentricity evolution of the four planets from the same simulation discussed in the previous section, focusing on a 10,000 yr window centered at approximately 100 Myr after the onset of evolution. This time interval is shown as a zoom-in to resolve the secular oscillations in eccentricity, which occur on relatively short timescales of hundreds to thousands of years. Although the eccentricities undergo continuous time-dependent variations driven by secular planet–planet interactions, their amplitudes remain modest throughout the interval, with all planets maintaining eccentricities below $e \lesssim 0.1$.}
While this value is {slightly} larger than the typical eccentricity for similar planets \citep[the average eccentricity being $e\sim 0.05\pm0.01$;][]{Gilbert2025}, to determine whether eccentricity oscillations of this magnitude will affect the prospects of the planets being habitable, we must consider how variations in eccentricity will affect the planet's irradiation.

We can use Equation 1 from \citep{Kane2016} to compute the flux on the planet's surface from stellar radiation: 
\begin{equation}
\frac{F_p}{F_\oplus} = \left( \frac{R_\star}{R_\odot} \right)^2 \left( \frac{T_{\text{eff}}}{T_{\text{eff},\odot}} \right)^4 \left( \frac{1\,\text{AU}}{a} \right)^2
\label{eq:radiation}
\end{equation}
where $F_p$ denotes the planet's flux, $F_{\oplus}$ the flux on Earth's surface, 
$R_\star$ and $R_\odot$ are the radii of the star and the Sun, respectively. $T_{\text{eff}}$ and $T_{\text{eff},\odot}$ are the effective temperatures of the star hosting the planets and the Sun respectively, and $a$ is the semi-major axis of the planet's orbit.

\begin{figure}
    \centering
    \includegraphics[width=1\linewidth]{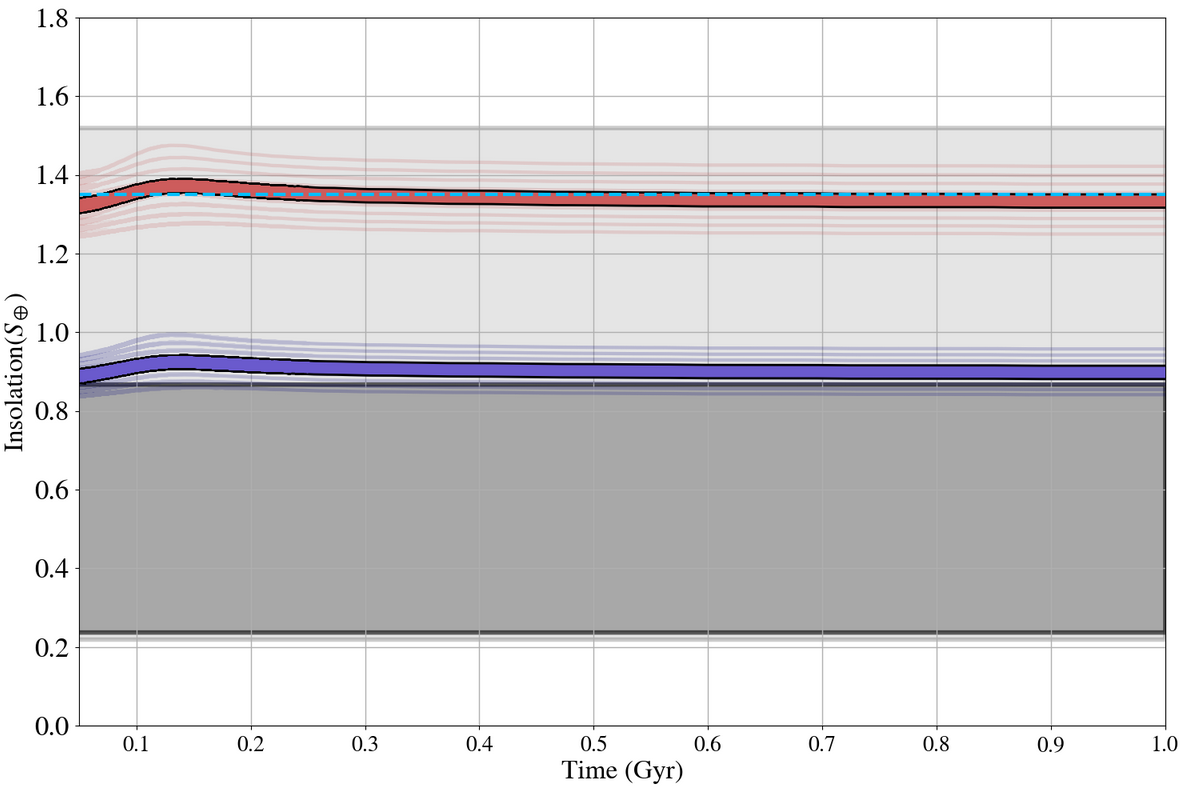}
    \caption{The irradiation (in units of Earth insolation, {solirad}) ($S_{\oplus}$) over time on planets TOI-700e (red lines) and TOI-700d (blue lines). The dark grey region is the conservative habitable zone, where the light grey region is the optimistic habitable zone {as found in the 1D modeling in \citep{Kopparapu2013}}. The bold lines represents the average of all of the simulations for each planet. {The dotted blue line represents to the location of the tidally locked habitable zone boundary, which comes from the 3D modeling of \citep{Kopparapu2017}. Using older 1D models,} both planets reside solidly in the potential habitable zone, and the evolution of their surface fluxes due to eccentricity evolution and stellar evolution do not compromise that. {However, TOI-700e is likely on the boundary of the tidally locked habitable zone, making its habitability prospects less clear. We assume the planets are at the equilibrium state found in Figure \ref{fig:obl_evolution} by 0.05 Gyr.} }
    \label{fig:insolation}
\end{figure}

To account for the influence of orbital eccentricity, which modifies the planet’s star–planet distance over time and thus the mean stellar irradiation, we can can scale Equation \ref{eq:radiation} by a factor of $\sqrt{1-e^2}$ \citep{Adams2006ecc,Bolmont2016,Gallo2024}:
\begin{equation}
\begin{split}
F_p(t) &= \dfrac{F_p(t)}{\sqrt{1 - e^2}} \\
&= \left( \frac{1}{\sqrt{1 - e^2}} \right) \left( \frac{R_\star}{R_\odot} \right)^2 \left( \frac{T_{\text{eff}}}{T_{\text{eff},\odot}} \right)^4 \left( \frac{1\,\text{AU}}{a} \right)^2,
\end{split}
\label{eq:radiation_updated}
\end{equation}
which describes a planet's eccentricity-corrected flux in units of {solirad} \citep{Mamajek2025}. 
In addition to secular eccentricity variations, changes in stellar luminosity can also alter the planet's flux. 
To compute the planet flux over time including both the evolving stellar luminosity and secular eccentricity evolution, we run 20 simulations for 1 Gyr each with the VPLanet \texttt{stellar}, \texttt{distorb} and \texttt{distrot} modules, {assuming the planets have already settled into the equilibrium state shown in Figure \ref{fig:obl_evolution} where the obliquities are 0 and the planets are tidally locked}. 
Each of the 20 simulations has initial parameters (including the planet semi-major axis $a$, stellar mass $M_{*}$) drawn uniformly from the 1$\sigma$ range reported in Table \ref{tab:bigtable}. The results are plotted in Figure \ref{fig:insolation}. Each individual line shows the results of one simulation, and the bold lines show the averaged evolution. The width of each individual line is due to the secular oscillations in eccentricity for each planet. 

We overlay on Figure \ref{fig:insolation} the conservative (dark grey) and optimistic (light grey) habitable zones \citep{Kopparapu2013}. The conservative habitable zone is defined as the region of space where water is likely to exist in liquid form{, assuming earth like parameters.} 
{The optimistic habitable zone is the region of space where water is able to exist assuming optimal conditions. The optimistic habitable zone (sometimes called Recent Venus Limit) derives from the empirical observation that Venus has no liquid water on its surface, but may have had oceans billions of years ago \citep{Kopparapu2013}.}

The irradiation levels for both planets are fairly even over time, as the eccentricity evolution does not significantly contribute to varying the planet's flux.
Similarly, the evolution at the start of the integration when the star is young has minimal effects on the characterization of either planet as habitable or not. 
Here, we can see that planet TOI-700e is solidly within the optimistic habitable zone and TOI-700d is on the edge between conservative and optimistic habitable zone.

{However, the claim that the definition of the optimistic habitable zone was based on (that Venus had early oceans) has recently been disputed \citep{Krissansen2021, Turbet2021, Constantinou2025}, as evidence now suggests that Venus was always dry. As a result, it is likely the optimistic habitable zone, which allows TOI-700e to be robustly habitable, is too permissive. In that case, a more reasonable inner limit of the habitable zone would likely be set by the tidal-lock limit described by \citet{Kopparapu2017}, located at 1.30 $S_\oplus$ for a star of TOI-700's effective temperature. If this inner boundary is adopted, TOI-700d remains securely within the habitable zone, whereas TOI-700e may lie near the edge of habitability, with portions of its posterior distribution falling {in the habitable zone and other parts residing too close to the star to be habitable.}}

\subsection{Water Retention}
\label{subsec:water}
\begin{figure}[htbp]
\centering
\includegraphics[width=\linewidth]{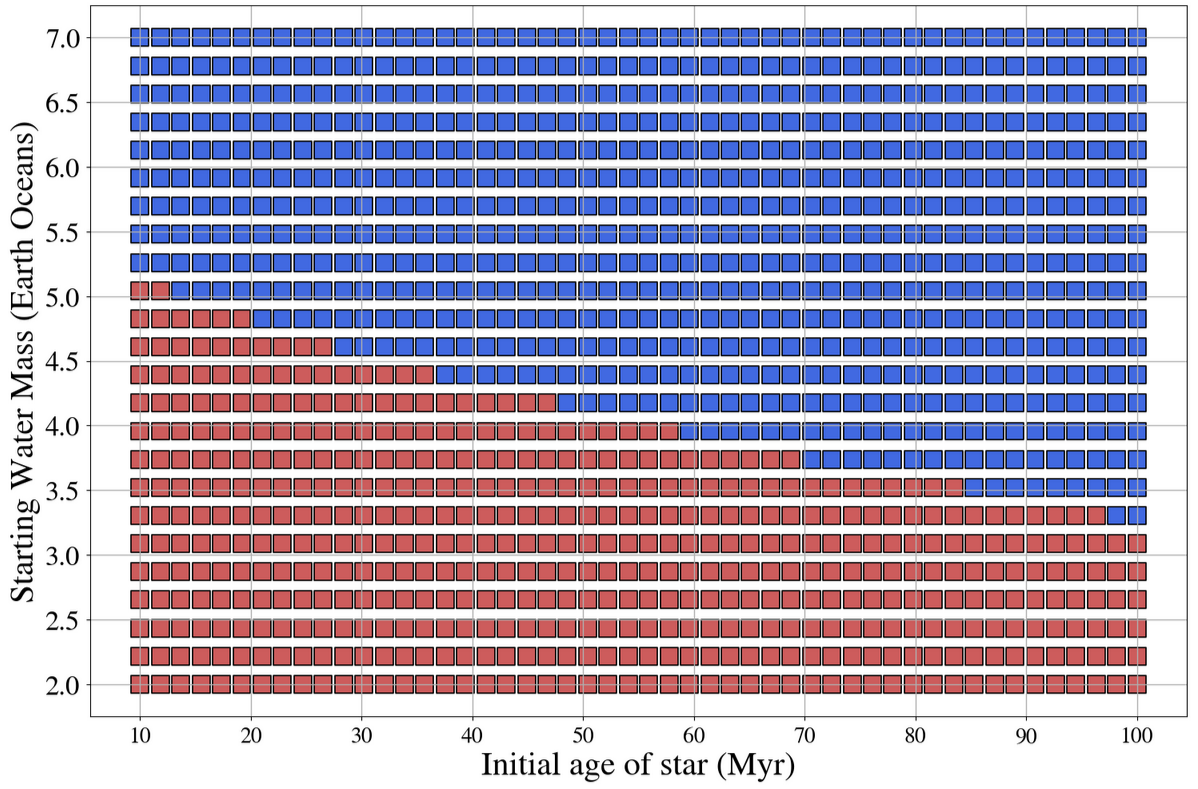}

\vspace{0.5em}

\includegraphics[width=\linewidth]{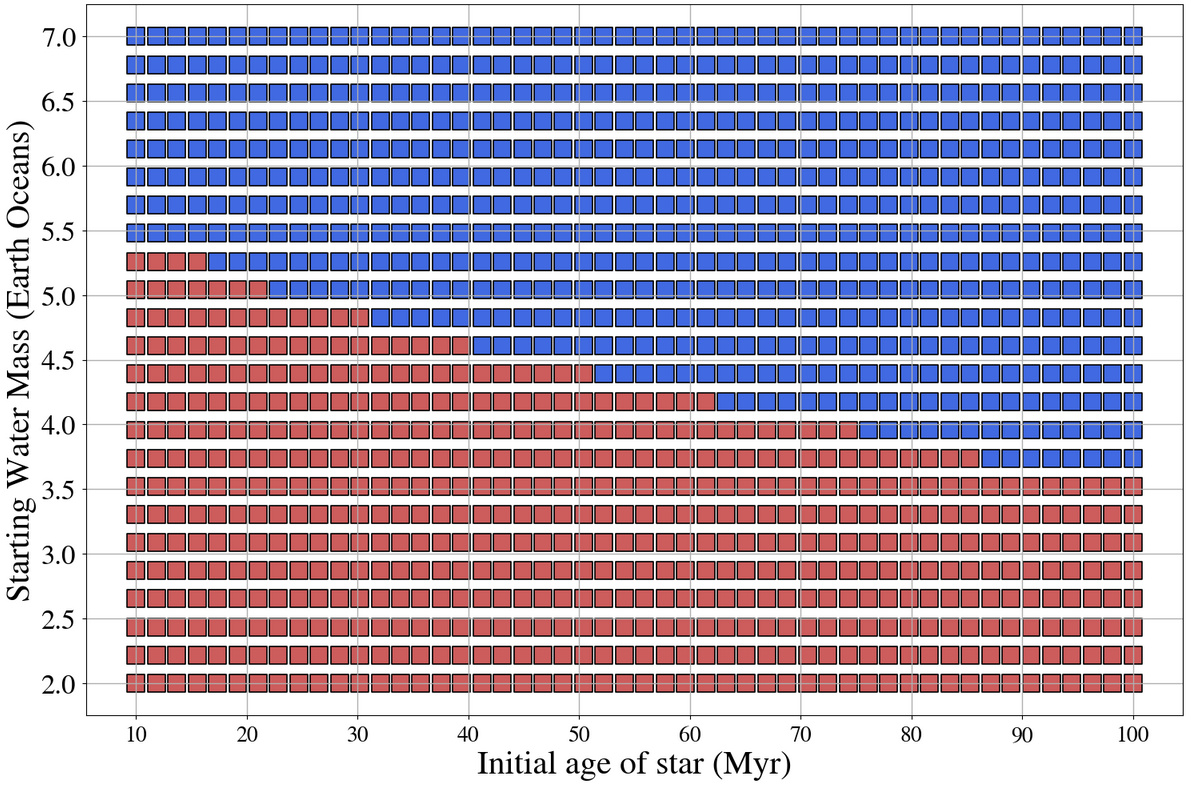}
\caption{Dependence of planetary water retention on the star's initial age (assumed to be the age of ocean formation on the planet) and starting water inventory. Each square corresponds to an independent simulation. Red squares indicate complete desiccation within 1 Gyr, while blue squares indicate retention of at least some initial water. The top panel shows results for TOI-700d, and the bottom panel for TOI-700e.}
\label{fig:evaporation}
\end{figure}

In the previous section, we considered whether the surface flux of TOI-700d and e places them in the habitable zone over $\sim$Gyr timescales. In this section, we will consider whether and under what initial conditions surface water may be maintained on these planets over the same timescale.

{We model this potential mass loss using the \texttt{AtmEsc} module of \texttt{VPLanet}, assuming that the dominant mode of mass loss will be energy-limited photoevaporation of hydrogen in the atmosphere \citep{Luger2015}. This will occur if there is some partial evaporation of surface water into the atmosphere. }

{The mass loss rate of hydrogen is computed by the energy-limited photoevaporation formula \citep{Watson1981, Lopez2012}:}
\begin{equation}
\frac{d {m_{\rm{EL}}}}{dt} = \epsilon \left( \frac{R_{\mathrm{XUV}}}{R_p} \right)^{2}
\left( \frac{3 F_{\mathrm{XUV}}}{4 G \rho} \right),
\label{eq:mdot}
\end{equation}
{where $\epsilon$ is the efficiency, $R_{\mathrm{XUV}}$ is the planet's radius in the XUV, ${R_p}$ its physical radius, $F_{\mathrm{XUV}}$ the incident stellar XUV flux, $G$ the gravitational constant, and $\rho$ the planetary bulk density. We assume that the planets, being mainly rocky (except for possibly TOI-700 c), will have $R_{\mathrm{XUV}}=R_p$.  }
{The mass loss of surface water can then be computed as \citep{Luger2015}:}
\begin{equation}
    \dot{m}_{ocean} = \frac{d {m_{\rm{EL}}}}{dt} \left( \frac{9}{1 + 8 \eta} \right)
    \label{eq:oceanloss}
\end{equation}
{when $\eta$ is a parameter that describes the ratio of escaping flux of oxygen to hydrogen, $F_{O} = \eta F_{H} / 2$.}
In \texttt{VPLanet}, surface water mass loss is assumed to occur {following the above prescription, driven by the} stellar XUV flux. 
Since water loss will thus be more efficient for young stars with higher XUV luminosities, a planet that forms its ocean earlier will experience a longer phase of increased XUV luminosity, which may cause it to lose more of its surface water via photoevaporation \citep{Borum2024}. {As a result, }in this suite of simulations, we keep all planetary and stellar parameters fixed to the best-fit values except for two: the starting water mass on each planet, and the initial age of the star. The initial age of the star is a proxy for the age at which the water formed on the planets{: if} water is primordial (on the planet's surface when it forms), then water loss will proceed when the protoplanetary disk dissipates. However, if the water is delivered to the surface later (either via outgassing from the planetary interior, \citealt{ElkinsTanton2011}, or if it is delivered by comets, \citealt{Albarede2009}), the photoevaporation may not begin until as late as 100 Myr or so.


In this set of simulations, we used the same parameters for the planet properties that were used in previous section, found in Table \ref{tab:bigtable}. 
Other than the initial water amount, we assumed Earth-like conditions for the other \texttt{atmesc} parameters \citep[following][]{Barnes2020}. 
The initial water mass was varied from 2.0 - 7.0 Earth oceans. 
The results of these simulations can be found in Figure \ref{fig:evaporation}. The top panel shows whether water remains after 1 Gyr of evolution for TOI-700d, and the bottom panel shows the same result for TOI-700e.
For both planets, low starting water and young initial star age leads to desiccation, whereas higher starting water and an older start lead to more water-rich planets. 
For TOI-700d (top panel of Figure \ref{fig:evaporation}), if the initial water inventory exceeds 5 Earth oceans, the planet will always retain its water. Below this threshold, retention depends on when the ocean formed relative to the host star’s age. The same is true for TOI-700e (Figure \ref{fig:evaporation}). Thus, if both planets began with sufficient initial water, they could retain their oceans and maintain potentially habitable conditions.
Our parameter sweep demonstrates that both TOI-700d and e can retain surface water, provided they start with a sufficient initial supply.

\label{subsec:ice}

\section{Discussion}
\label{sec:discuss}

{Our simulations indicate that dynamical interactions in the TOI-700 system are not, by themselves, a barrier to long-term habitability for either TOI-700d or TOI-700e. Across a range of initial conditions and observationally allowed system parameters, the system reaches an equilibrium spin state with tidal locking and zero obliquities (Figure \ref{fig:obl_evolution}), eccentricities remain low ($e \lesssim 0.1$; Figure \ref{fig:ecc_e_d}), and the resulting secular variations in orbit-averaged insolation are small compared to the separation between habitable-zone boundaries (Figure \ref{fig:insolation}). 
Taken together, these results suggest that the dominant uncertainties in the habitability prospects of TOI-700d and TOI-700e are not set by dynamical stability or secular interactions between planets, but instead by atmospheric state, initial water budget, and the XUV irradiation history (Figure \ref{fig:evaporation}}).

Our rotational evolution calculations in Section \ref{sec:obl} show that, for a reasonable range of planet parameters (including $Q$ and $k_2$), all four known planets in the system should tidally lock relatively early (within 10 Myr), with the outer potentially habitable planet TOI-700d synchronizing within $\lesssim 100$ Myr even for the smallest reasonable amount of tidal dissipation. This climate state will strongly and significantly influence surface water stability \citep[e.g.,][]{Kasting1993, Kopparapu2017, Shields2016}. Our results therefore motivate follow-up climate modeling that explicitly accounts for the rotation states of the TOI-700 planets. These analyses likely do not need to be coupled, because the results of this work indicate that planet–planet interactions are not a dominant driver of climate-relevant variability in this system. By the present day, obliquities are expected to be fully damped, and the planets have low eccentricities.


For both planets, an intrinsic planetary magnetic field may help the planets to retain their atmospheres \citep{Dong2020, Nishioka2023}. {Conversely, it may also facilitate atmospheric escape through processes like ion-pickup \citep{Gronoff2021}.} The local space weather environments {(such as solar wind) }may also play a role \citep{Cohen2020}. 

{Finally,} in our work, we find that TOI-700d can maintain surface water only if it started with a particularly large initial amount of water.
Previous analysis with different methods have found similar results \citep[e.g.,][]{Engelmann2021,Bonney2022}. 
In the simulations of \citet{Bonney2022}, about half of their models showed that TOI-700d loses all its surface water, and half their models allowed some water to be retained. 
Both our result and their results highlight the importance of well-determined planet parameters - the planet and star parameters are not yet well determined enough to determine with certainty whether TOI-700e and d are likely to have surface water.

\subsection{Caveats and Future Work}

Given the uncertainties in stellar and planetary properties, {the posterior distribution for TOI-700e is wide enough that  spans both regions of the habitable zone and straddles the \citet{Kopparapu2017} runaway greenhouse boundary,} meaning that some of the posterior is habitable for most atmospheric compositions and some of the posterior would require finely tuned atmospheric parameters to be habitable.
Improved orbital determinations with future observations will help narrow the posteriors and provide a clearer picture of the possible {habitability prospects} of TOI-700d and e.
However, not all parameters were varied in our analysis. For example, we fixed the orbital configurations of the two interior planets, TOI-700b and TOI-700c, to their best-fit values. Adding in variations over these planets' parameters would likely further broaden the possible results of our simulations. 

{In this work, we also assume constant tidal quality factors and Love numbers, whereas real planets may exhibit time-varying and frequency-dependent dissipation Similarly, our secular dynamics treatment captures long-term interactions between the known planets, but does not account for the effect of additional undiscovered planets. Since TOI-700e itself was an undiscovered planet after the discovery of the rest of the system in \citet{Gilbert2020} until the discovery of TOI-700e in \citet{Gilbert2023}, it is very likely that there are additional unknown system components. Finally, our chosen irradiation-based habitable-zone boundaries are approximate, as the planetary albedo and atmosphere composition can significantly change their locations. {Our assumed outer boundary of the habitable zone is based on 1D modeling, which will not be as accurate as 3D modeling \citep{Leconte2013}.} As a result, the most robust interpretation of our results is comparative: within the current uncertainties, TOI-700d is a stronger candidate for habitability, while TOI-700e has plausible but more threshold-sensitive habitability prospects.} {In particular, as shown in Figure \ref{fig:insolation}, TOI-700e resides on the edge of the tidally locked habitable zone from \citet{Kopparapu2017}, and depending on where its parameters fall in the posteriors it may either maintain habitable conditions or be too hot. }
{Future work aimed at more fully characterizing the 3D climate states of these planets, particularly the distribution and stability of surface ice coverage, would be valuable for assessing their habitability prospects.}

\section{Conclusion}
\label{sec:conclude}
Our simulations indicate that both TOI-700d and TOI-700e remain {plausible} candidates for habitability. We find that dynamical interactions within the compact TOI-700 system do not appear to destabilize their orbits or drive either of the planets outside of the habitable zone, and both planets maintain irradiation levels consistent with hosting liquid water. {All four planets in the system are likely in low-obliquity orbits, with spin periods equal to their orbital periods, and likely have low ($e<0.1$) orbital eccentricities.}
Our simulations of water retention show that each planet can retain surface water, provided a sufficient initial supply and favorable timing of ocean formation. 
Taken together, these results suggest that while the two planets may experience very different surface conditions, both remain viable targets for future studies of planetary habitability.

\section*{Acknowledgments}
We thank the referee for their extremely helpful comments.
This research has made use of the NASA Exoplanet Archive, which is operated by the California Institute of Technology, under contract with the National Aeronautics and Space Administration under the Exoplanet Exploration Program. This research has made use of NASA's Astrophysics Data System Bibliographic Services.

\software{\texttt{VPLanet} \citep{Barnes2020}, \texttt{matplotlib} \citep{Hunter:2007},
\texttt{pandas} \citep{mckinney-proc-scipy-2010, the_pandas_development_team_2024_13819579}, Astropy \citep{astropy:2013, astropy:2018, astropy:2022}, Jupyter \citep{kluyver2016jupyter}}

\bibliography{refs}{}
\bibliographystyle{aasjournal}
\end{document}